\documentclass[prl,aps,twocolumn,amsfonts,showpacs]{revtex4-1}
\usepackage{amsmath}
\usepackage{amssymb} 
\usepackage{color} 
\usepackage{graphicx}
\usepackage{hyperref}

\begin{document}

\title{Collective Dynamics of Bose--Einstein Condensates in Optical
  Cavities} 
\author{J. Keeling} 
\affiliation{University of Cambridge,
  Cavendish Laboratory, Cambridge, CB3 0HE, UK.}
\author{M. J. Bhaseen} 
\affiliation{University of Cambridge, Cavendish
  Laboratory, Cambridge, CB3 0HE, UK.}  
\author{B. D. Simons}
\affiliation{University of Cambridge, Cavendish Laboratory, Cambridge,
  CB3 0HE, UK.}  
\date{\today}
  
\begin{abstract}
  Recent experiments on Bose--Einstein condensates in optical cavities
  have observed a coherent state of the matter--light system ---
  superradiance.  The nature of these experiments demands
  consideration of collective dynamics.  Including cavity leakage and
  the back-reaction of the cavity field on the condensate, we find a
  rich phase diagram including a variety of multi-phase co-existence
  regions, and a regime of persistent optomechanical oscillations.
  Proximity to some of the phase boundaries results in critical
  slowing down of the decay of many-body oscillations, which can be
  enhanced by large cavity loss.
\end{abstract}


\maketitle

The tremendous advances in preparing Bose--Einstein condensates (BEC)
in optical cavities have opened new frontiers combining cold atoms and
quantum optics. It is now possible to enter the strongly coupled
regime of cavity quantum electrodynamics (QED)
\cite{Brennecke:Cavity,Colombe:Strong} in which atoms exchange photons
many times before spontaneous emission and cavity losses set in.  The
inherent cavity leakage also provides a valuable window on these
quantum many-body systems. In particular, it allows for {\em in situ}
non-demolition measurements of condensate properties via optical
transmission \cite{Ritsch:Probing,Chen:Numstat}.  The strong
matter--light coupling also supports collective dynamics and
back-reaction effects, stimulating new directions in cavity
optomechanics \cite{Brennecke:Opto,Ritter:Dyn} and self-organised
atomic ensembles
\cite{Domokos:Collective,Black:SO,Nagy:SO,Nagy:Self,Larson:MIRes,
  Larson:Stability,Gop:Emergent}.

\begin{figure}
\includegraphics[width=3.2in,clip]{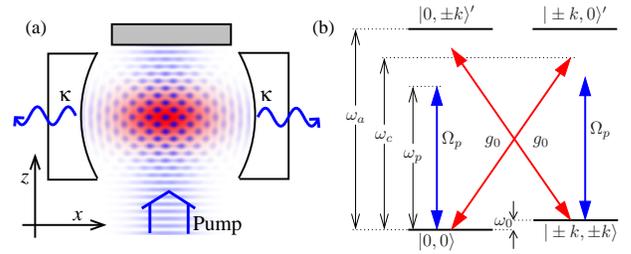}
\caption{Experimental setup \cite{Baumann:Dicke}. (a) BEC in a
  transversely pumped cavity  with pumping
  frequency, $\omega_p$, and strength $\Omega_p$, single-atom cavity
  coupling $g_0$, atomic transition frequency $\omega_a$, cavity
  frequency $\omega_c$, and cavity decay rate, $\kappa$. (b) Energy
  levels and pumping scheme showing the two-level splitting,
  $\omega_0=2\omega_r$, in the effective Dicke model, where
  $\omega_r=k^2/2m$ is the recoil energy.}
\label{Fig:BECDICKE}
\end{figure}

More recently, these capabilities have been elevated through
observation of the superradiance transition in BECs
\cite{Dimer:Proposed,Baumann:Dicke}.  The atom mediated coupling
between a transverse pump field and a cavity mode leads to a
realisation of the Dicke model
\cite{Dicke:Coherence,Hepp:Super,Wang:Dicke,Emary:Chaos}, in which
atomic momenta play the role of spin states; see
Fig.~\ref{Fig:BECDICKE}.  A significant merit of this approach is that
the energy splitting of the two-level systems is small enough so that
the Dicke superradiance transition may be realised with light at
optical frequencies \cite{Dimer:Proposed,Baumann:Dicke}.  These
experiments are a landmark in the study of quantum phase transitions
involving spins, and offer exciting and unique prospects for exploring
their static and dynamic properties.  Indeed, the time-dependent
nature of these measurements demands consideration of collective
dynamics.

Motivated by these developments we investigate the collective dynamics
of BECs in optical cavities. Our two primary goals are to establish
the generic behaviour, and to focus on the precise experimental
realisation in Ref.~\cite{Baumann:Dicke}.  We obtain a surprisingly
rich phase diagram for a broad range of parameters, and find distinct
regimes of dynamical behaviour, including several regions of
multi-phase co-existence, and regions of persistent optomechanical
oscillations.  For recent theoretical work see Ref.~\cite{Nagy:Dicke}.

The experiments in Ref.~\cite{Baumann:Dicke} consist of a $^{87}{\rm
  Rb}$ BEC with $N\sim 10^5$ atoms in an optical cavity with a
transverse pumping laser; see Fig.~\ref{Fig:BECDICKE}.  The excited
atoms may re-emit photons either along or transverse to the cavity
axis.  This process couples the zero momentum atomic ground state,
$|p_x,p_z\rangle=|0,0\rangle$, to the symmetric superpositions $|\pm
k,\pm k\rangle$, with an additional photon momentum along the cavity
or pump directions.  This yields an effective two-level system or
``spin'', where the splitting, $\omega_0$, is twice the atomic recoil
energy, $\omega_r=k^2/2m$.  One obtains an effective Dicke model for
collective spins, ${\bf S}$, of length $N/2$, coupled to radiation
$\psi$ \cite{Baumann:Dicke,Dimer:Proposed}
\begin{multline}
  H = \omega \psi^\dagger \psi + \omega_0S_z + US_z  \psi^\dagger \psi 
  + g
  (\psi^\dagger S^- + \psi S^+)
  \\+ g^\prime
  (\psi^\dagger S^+ + \psi S^-),
  \label{Dickeham}
\end{multline}
where, $\omega=\omega_c-\omega_p+NU_0(1+{\mathcal M})/2$,
$\omega_0=2\omega_r$, $U=U_0{\mathcal M}$, ${\mathcal M}$ is a matrix
element of order unity, and $U_0=g_0^2/(\omega_p-\omega_a)$ encodes
the back-reaction of the cavity light field on the BEC. The model
includes both co-rotating and counter-rotating matter--light
couplings, denoted $g$ and $g^\prime$. In the experiment
$g=g^\prime=g_0\Omega_p/(\omega_p-\omega_a)$ \cite{Baumann:Dicke}.

To describe the dynamics of the matter--light system (\ref{Dickeham})
we construct the Heisenberg equations of motion
\begin{equation}
\begin{aligned}
  \dot{S}^- &= 
  - i (\omega_0 + U \psi^\dagger \psi ) S^-
  + 2 i (g \psi + g^\prime \psi^\dagger) S_z,
  \\
  \dot{S}_z &=
  - i g \psi S^+ + i g \psi^\dagger S^-
  + i g^\prime \psi S^- - i g^\prime \psi^\dagger S^+,
  \\
  \dot{\psi} &= 
  - \left[ \kappa + i (\omega + U S_z) \right] \psi
  - i g S^- - i g^\prime S^+,
\end{aligned}
\label{eqmo}
\end{equation} 
where $S^\pm\equiv S_x\pm iS_y$, $\kappa$ is the cavity loss rate, and
we neglect atom loss \cite{Baumann:Dicke}.  Various limits of these
equations have been explored in different contexts.  For
$\kappa=g^\prime=0$ they describe fermionic pairing, where $\psi$ is
the Feshbach resonant closed state molecular field
\cite{Andreev:Noneqfesh,*Barankov:Coll}. This regime also arises for
polariton condensates and phase-locking of oscillators
\cite{Eastham:New,*Eastham:Phase}.  More recently, for $g=g^\prime$,
they have emerged in an elegant proposal for realising the Dicke model
\cite{Dimer:Proposed}.  As we will see, solutions of the more general
equations strongly influence $g=g^\prime$ dynamics.

\begin{figure}
\begin{center}
  \includegraphics[width=3.2in,clip]{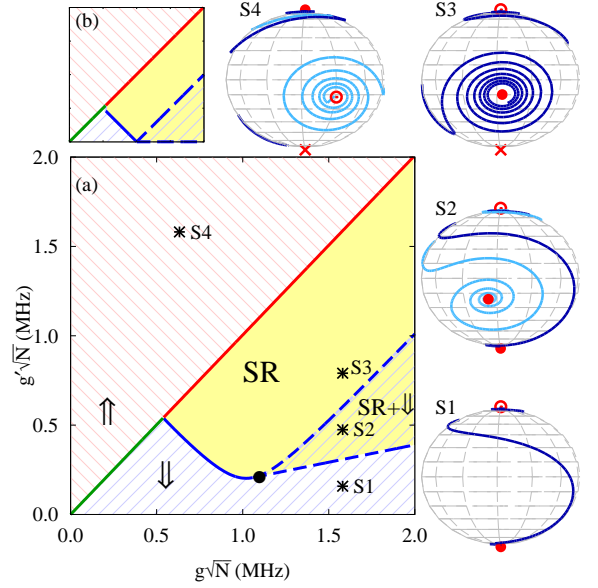}
  \caption{Dynamical phase diagram for $U=0$ and corresponding
      spin trajectories on the Bloch sphere. (a) Dynamical phase
    diagram for parameters $\omega=20$MHz, $\omega_0=0.05$MHz,
    $\kappa=8.1$MHz taken from Ref.~\cite{Baumann:Dicke}, showing the
    stable attractors of the nonlinear dynamics for $U=0$. The phases
    are: $\Downarrow$ all spins down $S_z=-N/2 $ and no photons,
    $\Uparrow$ all spins up $S_z=N/2$ and no photons, a non-trivial
    magnetised superradiant state with $\psi\neq 0$, and a co-existence
    region emanating from a tricritical point $\bullet$.  The
    separatrix $g^\prime/g=\sqrt{\beta_+/\beta_-}=1.0043$ is close to
    but distinct from unity. (b) Small cavity loss regime with
    $\kappa=1{\rm KHz}$ showing the evolution towards the
    superradiance transition at $g+g^\prime=\sqrt{\omega\omega_0}/N$
    in the equilibrium Dicke model
    \cite{Dicke:Coherence,Hepp:Super,Wang:Dicke,Emary:Chaos}.  The
    Bloch spheres S1-S4 show the stable $\bullet$, unstable $\circ$,
    and hyperbolic $\times$ fixed points (steady states) as well as
    characteristic trajectories in each of the phases. Examples of the
    time evolution for $g=g^\prime$ are given in
    Fig.~\ref{Fig:Timeevol}.}
\label{Fig:phase-boundary}
\end{center}
\end{figure}

In order to anchor the complete phase diagram, we start with $U=0$ and
consider $U\neq 0$ below.  Numerical solution of equations
(\ref{eqmo}), and the arguments below, yield the rich phase diagram in
Fig.~\ref{Fig:phase-boundary}, where the phases indicate stable
attractors of the long time dynamics.
Four distinct phases exist corresponding to all spins down and no
photons ($\Downarrow$), all spins up and no photons ($\Uparrow$), a
superradiant phase with photons (SR), and co-existence of the
supperradiant and down attractors; see S1-S4 in
Fig.~\ref{Fig:phase-boundary}.  Such co-existence, or bistability, is
related to the observed optomechanical oscillations in a different
system, where matter--light coupling is only through the back-reaction
\cite{Brennecke:Opto,Ritter:Dyn,Szirmai:Noise}.  In spite of the
cavity decay rate, $\kappa$, which may be large, the counter-rotating
terms stabilise superradiant steady states. Indeed, dropping the
derivatives in equation (\ref{eqmo}) renders algebraic equations, and
the determinantal condition for non-trivial solutions ($\psi\neq 0$)
yields
\begin{equation}
S_z =  
  \frac{-\omega \omega_0(g^2 + g^{\prime 2})\pm
    \sqrt{ (2 \omega\omega_0  g g^\prime)^2 - 
\omega_0^2 \kappa^2(g^2- g^{\prime 2})^2}}{2(g^2 - g^{\prime 2})^2}.
\label{szorderparam}
\end{equation}
The conditions for real physical solutions yield the blue phase
boundaries shown in Fig.~\ref{Fig:phase-boundary}\,(a).  Setting
$S_z=-N/2$ in equation (\ref{szorderparam}) yields the ``upper''
boundary shown in Fig.~\ref{Fig:phase-boundary}\,(a) \footnote{For
  $\kappa=0$, we recover Dicke model superradiance at
  $g+g^\prime=\sqrt{\omega\omega_0/N}$.  For $g=g^\prime$ it yields
  the results of Ref.~\cite{Dimer:Proposed}. See also
  Ref.~\cite{Nagy:SO}}.  The vanishing of the square root yields the
``lower'' boundary, $g^\prime=g\sqrt{\alpha_-/\alpha_+}$, where
$\alpha_\pm=\sqrt{\omega^2+\kappa^2}\pm\omega$, delineating the onset
of co-existence.  In order to identify the green phase boundary in
Fig.~\ref{Fig:phase-boundary}\,(a) it is necessary to consider the
stability of the steady states. We consider fluctuations about an
arbitrary configuration, ${\bf S}={\bf S}_0+\delta {\bf S}$,
$\psi=\psi_0+\delta\psi$, with frequency $\nu$.  Instability occurs if
${\rm Im}(\nu)>0$, and this yields the critical line
$g^\prime=g\sqrt{\beta_+/\beta_-}$ shown in
Fig.~\ref{Fig:phase-boundary}\,(a), where
$\beta_\pm=(\omega\pm\omega_0)^2+\kappa^2$.  This separates the stable
normal state $\Downarrow$ from the stable inverted state $\Uparrow$.
\begin{figure}
\includegraphics[width=3.2in,clip]{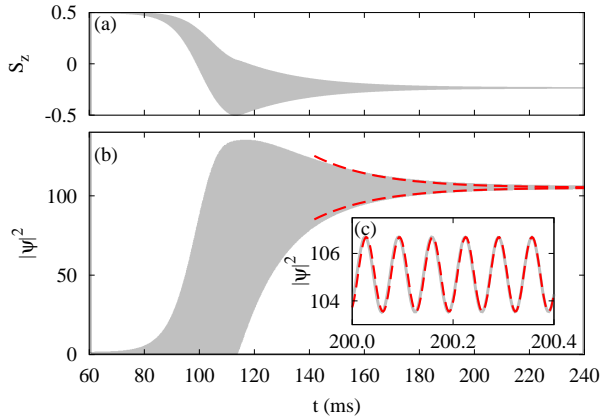}
\caption{Time evolution of the effective spin state and cavity
    photon population.  (a) Evolution of $S_z$ and (b) photon number
  for $g\sqrt{N}=g^\prime\sqrt{N}=0.791$MHz and $U=0$.  The long time behaviour shows
  the exponential envelope $|\psi|^2=|\psi_0|^2\pm {\mathcal A}e^{{\rm
      Im}(\nu)t}$ (dashed lines) where ${\mathcal A}$ is a
  non-universal amplitude dependant on the initial conditions, and the
  decay rate ${\rm Im}(\nu)=-\kappa\omega_0^2/(\kappa^2+\omega^2)$.
  (c) For $\omega_0\ll \omega$, the long time oscillation frequency is
  well described by the perturbative result ${\rm
    Re}(\nu)=\omega_0|S|/S_z+\delta$, where $\delta=4\omega
  g^2S_z^2/|S|(\kappa^2+\omega^2)$ is a small correction to leading
  term of order $\omega_0$. The short and intermediate time dynamics
  can be strongly affected by the existence of additional stable or
  unstable fixed points.}
\label{Fig:Timeevol}
\end{figure}
For the chosen parameters this gives $g^\prime/g=1.0043$, very close
to unity. The dynamics at $g=g^\prime$ may thus be strongly influenced
by proximity to this phase boundary.  The parameters used in
Fig.~\ref{Fig:phase-boundary} follow the hierarchy $\omega,\kappa\gg
g\sqrt{N}\gg \omega_0$, in which the photon decay rate,
$\kappa=8.1{\rm MHz}$, is much greater than the level spacing,
$\omega_0=0.046{\rm MHz}$ \cite{Baumann:Dicke}.  In this limit one
obtains a characteristic decay rate for the collective many-body
oscillations, ${\rm Im}(\nu)=-\kappa\omega_0^2/(\kappa^2+\omega^2)$,
as verified in Fig.~\ref{Fig:Timeevol} (b).  Notably, in the limit
$\kappa\rightarrow \infty$, corresponding to a {\em large} cavity loss
rate, this results in ${\rm Im}(\nu)\rightarrow 0$, or slow decay of
the collective oscillations.  This may be understood as critical
slowing down \cite{Hohenberg:TDCP}.  Further insight into this
$\kappa\rightarrow \infty$ dynamics may be gained by adiabatic
elimination of the fast photon field,
$\psi=-[i(g+g^\prime)S^x+(g-g^\prime)S^y]/(\kappa+i\omega)$, to derive
an effective equation of motion for the classical spins $\dot{\bf
  S}=\{{\bf S},H\}-\Gamma {\bf S}\times({\bf S}\times \hat z)$. Here
$H = \omega_0 S_z - \Lambda_+ S_x^2-\Lambda_-S_y^2$ is the
Lipkin--Meshkov--Glick Hamiltonian
\cite{Lipkin:LMG1,*Lipkin:LMG2,*Lipkin:LMG3,Morrison:Coll}, with
$\Lambda_\pm\equiv\frac{ \omega}{\kappa^2 + \omega^2} (g\pm
g^\prime)^2$ and $\Gamma\equiv\frac{2 \kappa}{\kappa^2 + \omega^2}
({g^\prime}^2-g^2)$.  The additional term takes the form of damping in
the Landau--Lifshitz--Gilbert equations
\cite{Landau:LLG,*Gilbert:LLG}.  Depending on the sign of $\Gamma$
this favours spin alignment either parallel or anti-parallel to the
$z$-axis.  The sign change at $g=g^\prime$ is consistent with the
$\kappa\rightarrow\infty$ limit of the phase boundary,
$g^\prime=g\sqrt{\beta_+/\beta_-}$, which separates the $\Downarrow$
and $\Uparrow$ steady states.  It is interesting to contrast the
emergence of integrable dynamics for $g=g^\prime$ and
$\kappa\rightarrow \infty$, with the chaotic behaviour when
$g=g^\prime$ and $\kappa=0$ \cite{Emary:Chaos}. Moreover, for $g\neq
g^\prime$ the dynamics is non-Hamiltonian.

\begin{figure}
\includegraphics[width=3.2in,clip]{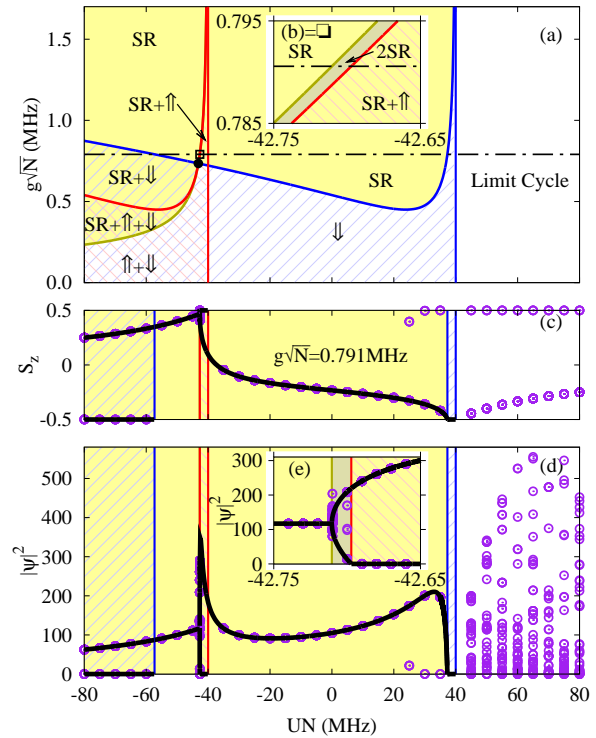}
\caption{Dynamical phase diagram as a function of $U$ and long time
  steady states. (a) Dynamical phase diagram of model (\ref{Dickeham})
  with $g=g^\prime$ and parameters $\omega=20$MHz, $\omega_0=0.05$MHz,
  $\kappa=8.1$MHz taken from Ref.~\cite{Baumann:Dicke}.  The blue, red
  and gold critical boundaries correspond to equations (\ref{cb}).
  (b) Magnified portion showing the appearance of a bistable
  superradiant phase (2SR) corresponding to co-existence of both roots
  of equation (\ref{sz1}).  (c,d) Trajectory along the dashed-dotted
  line showing the comparison between steady state solutions and
  numerical integration of the equations of motion at 360ms.  The
  region to the right of the blue asymptote corresponds to a limit
  cycle.  For each value of $U$ we take a variety of initial
  conditions with $\psi=1$ and ${\bf S}$ uniformly distributed over
  the Bloch sphere.  (e) Magnified portion of the bistable
  superradiant phase (2SR) showing the agreement between the steady
  states and numerical integration.  }
\label{Fig:gvU}
\end{figure}

Having discussed the dynamics of the model (\ref{Dickeham}) for $U=0$,
let us now consider $U\neq 0$. In order to make close contact with the
experimental realisation in Ref.~\cite{Baumann:Dicke} we henceforth
set $g=g^\prime$.  In Fig.~\ref{Fig:gvU}\,(a) we present the dynamical
phase diagram as a function of $U$.
The entire topology may gleaned analytically from the steady state
solutions of equation (\ref{eqmo}).  These reveal two classes of
superradiant solutions incorporating both $U$ and $\kappa$.  The first
class has a non-vanishing photon population
\begin{equation}
|\psi|^2=
  \frac{ 4 g^2 }{\tilde{\omega}^2 + \kappa^2}
  \left( \frac{N^2}{4} - S_z^2 \right),
\end{equation}
where $\tilde\omega\equiv \omega+US_z$, and
\begin{equation}
  S_z = - \frac{\omega}{U}
  \pm \sqrt{ \frac{
      g^2 ( 4\omega^2 - U^2 N^2) - U \omega_0 \kappa^2}{
      U^2 (\omega_0 U + 4 g^2)}}, \quad S_y=0,
\label{sz1}
\end{equation}
and $S_x$ is determined by the normalisation of the spins.  Physical
solutions require $|S_z|\le N/2$.  In the limit $\kappa,U\rightarrow
0$ we recover the results of equilibrium superradiance
\cite{Dicke:Coherence,Hepp:Super,Wang:Dicke,Emary:Chaos}, and for
$U=0$ they reduce to those of Ref.~\cite{Dimer:Proposed}.  For
sufficiently large negative $U$ it is possible for equation
(\ref{sz1}) to develop unphysical complex roots. In this case one may
satisfy equations (\ref{eqmo}) with $\tilde\omega_0\equiv
\omega_0+U|\psi|^2=0$, $\tilde\omega\equiv \omega+US_z=0$, and
\begin{equation}
\psi = i \sqrt{\frac{-\omega_0}{U}},
\quad S_x = - \frac{\kappa}{2g} \sqrt{ \frac{-\omega_0}{U} }, \quad
  S_z = - \frac{\omega}{U},
\label{sr2}
\end{equation}
where $S_y$ is determined by normalisation.  Physical solutions have
$S_x^2+S_z^2\le N^2/4$.  In general these distinct solutions are
connected for $g\neq g^\prime$, so we do not distinguish them in
Fig.~\ref{Fig:gvU}\,(a).  Nonetheless, it is important to keep track
of them for analytical work when $g=g^\prime$.  Figure
\ref{Fig:gvU}\,(a) consists of three phase boundaries corresponding to
instability of $\Downarrow$ (blue), instability of $\Uparrow$ (red),
and existence of the second-type superradiant phase (gold):
\begin{equation}
g_{\Downarrow,\Uparrow} =
  \sqrt{
    \frac{\pm[(\omega\mp\omega_U)^2 + \kappa^2] \omega_0 U}{
      8 \omega_U(\omega\mp\omega_U)}}, \quad 
g_\ast = \frac{\kappa}{2} 
  \sqrt{ \frac{\omega_0 U}{\omega^2 - \omega_U^2}}, 
\label{cb}
\end{equation}
where $\omega_U \equiv UN/2$. Instability of the normal state
$g_\Downarrow$ has also been considered for thermal clouds in a ring
cavity \cite{Nagy:SO}.  The result for $g_\ast$ delimits the region,
both for equation (\ref{sz1}) and equation (\ref{sr2}), to have real,
physical solutions.  All three of these boundaries intersect at $U =
-2N^{-1}\sqrt{\omega^2 + \kappa^2}$, $g= \sqrt{-\omega_0 U/4}$, as
shown in Fig.~\ref{Fig:gvU}\,(a). Upon increasing $g$ one finds a
phase where two distinct first-type superradiant solutions co-exist;
see Fig.~\ref{Fig:gvU}\,(b).  This is borne out in
Fig.~\ref{Fig:gvU}\,(c,d,e), where we compare the steady states with
direct integration of equations (\ref{eqmo}) along the dashed-dotted
line in Fig.~\ref{Fig:gvU}\,(a).  We integrate over a period of 360 ms
to eliminate the transitory effects of critical slowing down discussed
earlier. In addition, Fig.~\ref{Fig:gvU} contains several regions
involving co-existence of superradiant {\em and} non-superradiant
phases.

The scattered points in Fig.~\ref{Fig:gvU}\,(d), beyond the
$\Downarrow$ boundary at $U=2\omega/N$, correspond to limit cycles
rather than steady states. Here $S_z=-\omega/U$ and $\psi$ is purely
imaginary. Writing $S^-=re^{-i\theta}$, where
$r=\sqrt{N^2/4-\omega^2/U^2}$, yields $\partial_t\theta = \omega_0 + U
|\psi|^2$, and $(\partial_t + \kappa) \psi = - 2igr\cos(\theta)$, with
limit cycle behaviour. For $\kappa\gg \omega_0+U|\psi|^2$, these
describe a damped driven pendulum.

\begin{figure}
\includegraphics[width=3.2in]{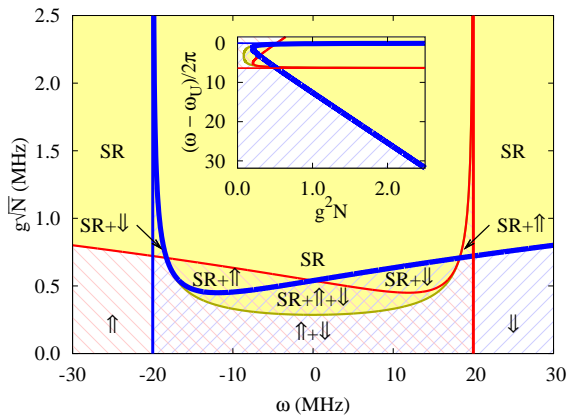}
\caption{Dynamical phase diagram as a function of $\omega$ for
    the experimental parameters used in Ref.~\cite{Baumann:Dicke}.
  Dynamical phase diagram in the $g\sqrt{N}$ versus $\omega$ plane for
  $UN=-40$MHz, where the blue, red and gold phase boundaries are given
  by equation (\ref{cb}) and correspond to those in
  Fig.~\ref{Fig:gvU}\,(a).  The thick blue line is the boundary of
  stability of the $\Downarrow$ state that would be seen on increasing
  $g$ as in Ref.~\cite{Baumann:Dicke}.  Inset: Phase diagram
  re-plotted as a function of $g^2N$ for comparison with 
Fig.~5 of Ref.~\cite{Baumann:Dicke}.}
\label{Fig:omg}
\end{figure}

Having confirmed the overall phase diagram in Fig.~\ref{Fig:gvU}\,(a)
as a function of $U$, let us finally focus on the specific value
$UN=-40$MHz used in Ref.~\cite{Baumann:Dicke}. In Fig.~\ref{Fig:omg}
we plot the phase diagram as a function of $\omega$ for this fixed
value of $U$. We see that the superradiance boundary is accompanied by
several regions of multi-phase co-existence.  It would extremely
interesting to investigate this experimentally. The inset shows the
same data shifted and rescaled for comparison with Fig. 5 of
Ref.~\cite{Baumann:Dicke}.

In summary, we have discussed the collective dynamics of BECs in
optical cavities. We obtain a rich phase diagram with different
regimes of dynamical behaviour, including several regions of
multi-phase co-existence and the slow decay of many-body oscillations.
Amongst our findings is a regime of persistent optomechanical
oscillations described by a damped driven pendulum.  Given the strong
interest in cavity optomechanics \cite{Brennecke:Opto,Ritter:Dyn} this
may be a profitable region to explore experimentally.  Further
directions include the impact of cavity axis pumping
\cite{Tomadin:Noneq} and photon correlations.  Experiments in which
the coupling $g$ is quenched through the phase boundaries may help
explore this rich dynamics.

We are extremely grateful to K. Baumann, F. Brennecke, T.  Esslinger
and M. K\"ohl for illuminating discussions.  MJB and JK acknowledge
ETH Z\"urich, G. Blatter, S. Schmidt, and H. T\"ureci for hospitality
and interactions. MJB and BDS acknowledge EPSRC grant
no. EP/E018130/1. JK acknowledges EPSRC grant no. EP/G004714/1.

\end{document}